\documentclass[doublecol]{epl2} 
\usepackage{ogonek}

\def\(({\left(}
\def\)){\right)}                       
\def\[[{\left[}
\def\]]{\right]}

\newcommand{\be}{\begin{equation}}
\newcommand{\ee}{\end{equation}}
\newcommand{\bea}{\begin{eqnarray}}
\newcommand{\eea}{\end{eqnarray}}

\title{Potts Glass on Random Graphs}

\author{Florent Krz\k{a}ka{\l}a\inst{1} \and Lenka Zdeborov\'a\inst{2}
\inst{3}
}
\shortauthor{Florent Krz\k{a}ka{\l}a and Lenka Zdeborov\'a}

\institute{
  \inst{1}  PCT, UMR 7083 CNRS-ESPCI, 10 rue Vauquelin, 75231 Paris, France \\
\inst{2} Universit\'e Paris-Sud, LPTMS, UMR8626,  B\^{a}t. 100, Universit\'e Paris-Sud 91405 Orsay cedex, France\\
\inst{3} CNRS, LPTMS, UMR8626, B\^{a}t. 100, Universit\'e Paris-Sud 91405 Orsay cedex, France
}

\abstract{We solve the $q$-state Potts model with anti-ferromagnetic
  interactions on large random lattices of finite coordination.  Due
  to the frustration induced by the large loops and to the local
  tree-like structure of the lattice this model behaves as a mean
  field spin glass. We use the cavity method to compute the
  temperature-coordination phase diagram and to determine the location
  of the dynamic and static glass transitions, and of the Gardner
  instability.  We show that for $q\ge 4$ the model possesses a
  phenomenology similar to the one observed in structural glasses.  We
  also illustrate the links between the positive and the
  zero-temperature cavity approaches, and discuss the consequences for
  the coloring of random graphs.  In particular we argue that in the
  colorable region the one-step replica symmetry breaking solution is
  stable towards more steps of replica symmetry breaking.
}


\begin{document}

\maketitle

There are mainly two types of lattices for which the mean field
theory of spin glasses is exact. The first type is the fully connected
lattice underlying the canonical Sherrington--Kirkpatrick
model~\cite{SK}. Parisi introduced the replica symmetry breaking (RSB)
scheme \cite{BOOK} to obtain a solution of this model that was much
later proven to be rigorously exact~\cite{Talagrand}.  Similar
solutions have been derived for the p-spin model \cite{Pspin} and for
the Potts glass model \cite{Sompolinsky}, and these played a major
role in the development of the mean field theory of the structural
glass transition \cite{KirkpatrickThirumalai,BiroliMezard}.

The second type of lattice for which the mean field
theory is exact is given by large random graph with fixed average
degree (or connectivity), a case commonly refereed to as ``Bethe
lattice'' in the physics literature~\cite{VianaBray}.  These lattices are
considered as more realistic since the notions of
distance and neighboring can be naturally defined. Few years ago,
M\'ezard and Parisi \cite{MP99,MP03}, using the cavity method, have
adapted the replica symmetry breaking scheme to solve models on such sparse
lattices. The theoretical activity following this breakthrough was
mainly concentrated on the zero temperature limit \cite{MP03} of
various models on random graphs, motivated by the equivalence with the
random cases of hard combinatorial optimization problems such as
graphs coloring or the satisfiability of boolean formulae
\cite{Science}.  The last few years have indeed witnessed spectacular
successes in this direction\cite{Science}.

In this paper, we consider a $q$-state Potts model with
anti-ferromagnetic nearest neighbors interactions on random
graphs. Due to the frustration induced by large loops and to the
locally tree-like structure of the graph, the model actually behaves
as a mean field spin glass. Using the cavity approach, we study
systematically the phase diagram for different temperatures $T$,
average connectivities $c$, and number of states (colors) $q$.
Determining the different phases, and which step of replica symmetry
breaking is needed for given $T,c,q$ are the main contributions of
this work. The zero temperature limit of this model maps to the
$q$-coloring problem, and has been studied extensively
\cite{ColoringSaad,Coloring,ColoringFlo,ColoringMarc,MM05,US1,US2},
while the positive temperature studies are limited to
\cite{ColoringSaad,MM05,US2}. Our results give a generic picture of
Potts glasses on sparse random graphs, and have interesting
consequences for constraint satisfaction problems as well. Finally, we
also discuss the subtle connections between the positive \cite{MP99}
and zero temperature \cite{MP03} cavity formalisms, and correct some
confusions that have been made in the literature in the use of these
two approaches.

\section{Anti-ferromagnetic Potts glass}
\label{sec:Model}
We consider a $q$-state anti-ferromagnetic Potts model. Given a graph $G =
({\cal V,E})$ defined by its vertices ${\cal V}=\{1,...,N\}$ and edges
$(i,j)\in {\cal E}$ the Hamiltonian reads
\be
{\cal H} = \sum_{(i,j) \in {\cal E}} \delta(\sigma_i,\sigma_j)\ ,
\label{Ham}
\ee
where $\sigma=1,\dots,q$ are the values of the different Potts spins. On a
bipartite graph (like a $d$-dimensional lattice), the ground-state is
anti-ferromagnetically ordered. On large random graphs, however, the
frustration induced by the large loops does not allow the anti-ferromagnetic
order and a glassy solution appears instead.

We will consider two ensembles of sparse random graphs: the Erd\H{o}s-R\'enyi
(ER) graphs with Poissonian degree distribution of mean $c$, and the regular
graphs with fixed degree $c$. Both are locally tree-like in the thermodynamic
limit so that the length of the shortest loop going through a random vertex
diverges as $\log{N}$ when $N \to \infty$ (our results generalize to other
locally tree-like ensembles).

It is interesting to discuss the relation of the model defined by the
Hamiltonian (\ref{Ham}) with the one defined by the following disordered
Hamiltonian
\be \hat{\cal H} = \sum_{i<j: (i,j) \in {\cal E}}
\delta\((\sigma_i,\pi_{ij}\((\sigma_j\))\))\, ,
\label{Ham2}
\ee
where $\pi_{ij}$ represents a random permutation of the $q$ colors. In this
disordered anti-ferromagnetic Potts model, the anti-ferromagnetic order is
{\it explicitly} destroyed by the disorder independently of the lattice.
Notice that a different disordered Potts model was defined originally
\cite{ElderfieldSherrington,Sompolinsky} but the advantage of our definition
(\ref{Ham2}) is that due to the existence of a gauge symmetry, the total
equilibrium magnetization is zero at all temperatures, independently of the
lattice.  Indeed, if we change the color of an arbitrary spin $i$, then there
always exists an energy conserving gauge transformation of the interactions
between spin $i$ and its neighbors such that the probability of the new set of
interactions is the same as of the original one. After averaging over disorder
the magnetization is zero, and self-averaging implies that it is also zero on
a single large system.  The model defined by (\ref{Ham2}) is thus convenient
for numerical simulations of the glassy phase, even on regular finite
dimensional lattices. Note that a ferromagnetic version of (\ref{Ham2}) was
already defined and studied in \cite{Enzo}.

The two models defined by (\ref{Ham}) and (\ref{Ham2}) share the same solution
on sparse random lattices. Indeed, by the gauge transformation, the two models
can be made equivalent on a large tree-like neighborhood of a vertex, the
difference being pushed on the boundary of this neighborhood.  Inside a pure
state the boundary conditions caused by the rest of the graph are uncorrelated
and color symmetric (a color asymmetric, ordered, solution is ruled out by the
frustrating long loops), thus a point-wise random permutation does not change
their statistical properties and the same solution is obtained for both the
models. We actually use this equivalence to resolve the numerical instability
towards the anti-ferromagnetic solution in the cavity iterative equations for 
the model (\ref{Ham}) by
randomly permuting the colors after each iteration, which is exactly what the Hamiltonian (\ref{Ham2}) would do.

\section{Cavity methodology reminder}
\label{sec:Method}
The cavity method \cite{MP99} allows to solve models (\ref{Ham},\ref{Ham2}) 
on large sparse random graphs. The results are expected to be exact (although a
rigorous proof is still missing) provided that the correct level of
replica symmetry breaking is considered.  A detailed derivation of the
equations for Hamiltonian (\ref{Ham}) is presented in \cite{US2},
we discuss only the parts necessary for the presentation of our results.

\subsection{The liquid phase}
The high-temperature liquid, or paramagnetic, phase is characterized by an
exponential decay of correlation functions and is associated with the
existence of a single pure state. The simplest version of the cavity method,
called replica symmetric (RS), describes correctly this phase
\cite{ColoringSaad,US2}. A first, simple, way to check the appearance of a
spin glass phase is to compute the temperature where the spin-glass
susceptibility diverges, or equivalently where the RS cavity equations do not
converge on a single graph or, in other words, where the RS solution is {\it
  locally unstable}. When this happens, a continuous phase transition towards
a spin glass phase arises. The temperature of the local RS instability is
given by \cite{US2}
\be T_{\rm local} (q,c) =  - 1/{\ln\(( 1-\frac q{\sqrt{\kappa} +1}\))} \, ,
\label{stab_KS}
\ee
where $\kappa=c$ for ER graph, and $c-1$ for regular graphs. 

It is however common to observe the discontinuous appearance of a spin
glass solution before the local instability sets in. In order to check
this point, it has been argued recently \cite{Glass, MM05} that the
decay of the so-called point-to-set correlations provides a {\it
  necessary and sufficient} condition for the validity of the RS
solution.  When the temperature is lowered, a dynamical temperature
$T_d$ might be reached below which the point-to-set correlation does
not decay to zero and the system undergoes a dynamical phase
transition, as the equilibration time of ``local flip'' Monte-Carlo
dynamics diverges \cite{Glass}. As shown in~\cite{MM05}, this
correlation does not decay to zero {\it if and only if} it exists a
non-trivial solution to the one-step replica symmetry breaking
equation when the so-called Parisi parameter $m=1$ (see below). 
Below $T_d$ the RS pure state splits into 
exponentially many disconnected pure states.
\begin{figure*}[!ht]
\begin{center} 
  \includegraphics[width=1.0\textwidth]{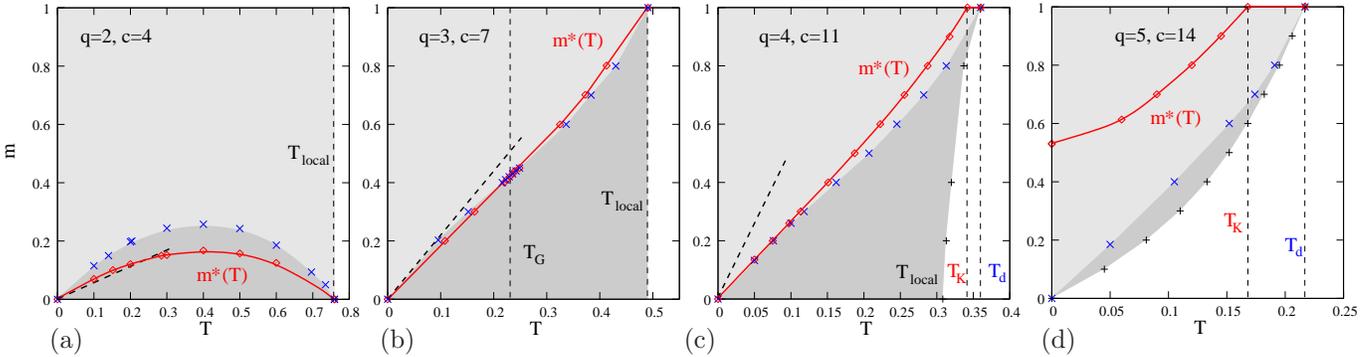}
  \put(-495,-5){(a)}
  \put(-370,-5){(b)}
  \put(-255,-5){(c)}
  \put(-120,-5){(d)}
  \caption{A set of m-T diagrams for the $q$-state anti-ferromagnetic
    Potts model on $c$-regular random graphs: a non-trivial solution
    of eq.~(\ref{1RSB}) exists in the shaded region. The curves with
    diamond data points $m^*(T)$ (red online) represent the
    thermodynamic value of the parameter $m$.  The darker region,
    delimited with crosses (blue online) is type II unstable towards
    further steps of replica symmetry breaking. The dotted lines
    (absent in (d)) represent the type I instability $m_I(T)$
    calculated analytically close to $T=0$, 1RSB is type I unstable
    for $m>m_I(T)$. $T_d$, $T_K$, $T_G$ and $T_{local}$ are the dynamical,
    Kauzmann, Gardner and local temperature (respectively).
    (a) is an example of continuous
    (Sherrington-Kirkpatrick-like) transition towards more than one
    step of RSB; (b) is an example of continuous transition to the
    1RSB phase, with a Gardner transition at lower temperatures; (c)
    is a discontinuous transition in the uncolorable phase and (d) is a 
    discontinuous transition in the colorable phase.}
  \label{FIG-mt} 
\end{center}
\end{figure*} 
\subsection{The glass phase with many states} The one-step replica symmetry
breaking (1RSB) cavity method deals with this situation. It allows to compute
the {\it complexity $\Sigma(f)$} defined as the logarithm of the number of
pure states of internal free energy density $f$. In order to do so, we 
follow~\cite{Remi} and define a potential $\Phi(\beta,m)$ as
\be e^{-\beta m N \Phi(\beta,m)} = \sum_{\alpha} e^{-\beta m N
  f_\alpha(\beta)} = e^{N[-\beta m f(\beta)+\Sigma(f)]},
   \label{Legendre}
\ee
where the sum is over all states $\alpha$, and
$m=\partial_f \Sigma(f) / \beta $ a temperature-like
parameter in the Legendre transform~\cite{Remi,MP99}. 
In the closely related replica formalism, the parameter $m$ defines the block size of Parisi's overlap matrix \cite{BOOK}.

For regular graphs
$\Phi(\beta,m)$ can be computed from the fixed point of the 1RSB equation
\begin{equation}
  P(\vec \psi) = \frac{1}{Z_1} \prod_{k=1}^{c-1} \int d \vec \psi^k  P(\vec \psi^k) \, \delta\left[\vec \psi- \vec {\cal F}(\{\vec\psi^k\}) \right] (Z_0)^m,\hspace{-0.09cm} \label{1RSB}
\end{equation}
where $P(\vec \psi)$ is the unknown probability distribution of $q$-component
normalized vectors $\vec \psi$, $Z_1$ a normalization, and the function
${\cal F}$ is defined 
component-wise as
\begin{equation}
  \psi_r = {\cal F}_r(\{\psi_r^k\})= \frac{1}{Z_0} \prod_{k=1}^{c-1} \left[1- \left(1-e^{-\beta}\right) \psi_r^{k} \right]\, , \label{update}
\end{equation}
where $r=1\dots q$,
$Z_0$ is a normalization constant and each component of the vector
$\vec \psi$ is the probability that a node, in absence of one of its neighbors,
takes the corresponding color \cite{MP99}.  For graphs with fluctuating degree
eq.~(\ref{1RSB}) has to be written edge-dependently and an average over the
edges of the graph has to be performed \cite{MP99,US2}. In both cases,
eq.~(\ref{1RSB}) can be solved numerically using the population dynamics method
\cite{MP99} where the probability distribution $P(\vec \psi)$ is represented
by a population of many elements $\vec \psi$.  At $m=1$, a formal substitution
introduced in \cite{MM05} (see also \cite{US2}) allows to rewrite the
averaging of eq.~(\ref{1RSB}) in a simple way for the fluctuating
connectivity graphs as well.

\section{The two-temperatures diagram}
The temperature $T$ ({\it resp.} temperature-like parameter $m$) in
(\ref{1RSB}) is related to the energy ({\it resp.} size) of the states.  It is
thus convenient to consider the solution of eq.~(\ref{1RSB}) in the $m-T$
diagram.  This representation moreover clarifies the connection of the positive
temperature approach with the different zero-temperature limits of
\cite{Coloring, ColoringFlo} and \cite{ColoringMarc,US2}. We plot several such
diagrams in fig.~\ref{FIG-mt}.  The region where a non-trivial (different
from the RS one) solution of (\ref{1RSB}) exists is shaded in light and darker
grey. At $m=1$ a non-trivial solutions exists for all $T \leq T_d$.  At $m=0$
the absence of the re-weighting term $Z_0^m$ makes the solution of
(\ref{1RSB}) be non-trivial if the RS equations do not converge on a single
graph, i.e., for all $T<T_{\rm local}$.

The next step is the computation of the thermodynamic value of the parameter
$m^*(T)$. The free energy in the 1RSB phase reads
\begin{equation}
  -\beta f_{\rm 1RSB}= \max_{f:\, \Sigma(f)\ge 0} \left[-\beta f + \Sigma(f)\right]. \label{max}
\end{equation}
Call $\tilde f$ the free energy for which $\partial_f \Sigma(\tilde f)=\beta$ (so that $m=1$, see eq.~(\ref{Legendre})). If
$\Sigma(\tilde f)\ge 0$ then $-\beta f_{\rm 1RSB} = -\beta \tilde f +
\Sigma(\tilde f) = -\beta \Phi(\beta,1)=-\beta f_{\rm RS}$ and the
free energy is equal to the RS one.  This explains the importance of
the value $m=1$: as long as the complexity $\Sigma(m=1)>0$, then
$m^*=1$. If $\Sigma(m=1) < 0$, however, the pure states corresponding
to the free energy $\tilde f$ are almost surely absent in the large
size limit and to maximize eq.~(\ref{max}) we choose instead $f^*$
such that $\Sigma(f^*)=0$. Consequently $f_{\rm 1RSB}=f^* > f_{\rm
  RS}$ and a genuine phase transition happens at the so-called
Kauzmann temperature $T_K$ where $\Sigma(m=1)=0$.

Between $T_d$ and $T_K$ the thermodynamics is dominated by an
exponential number of states and the total free energy is equal to
liquid one, so that the dynamical transition is not a true transition
in the Ehrenfest sense.  Below the Kauzmann temperature the
thermodynamics is dominated only by a finite number of them (although
exponentially many states still exist, they are thermodynamically
negligible).  At $T=T_K$, the free energy has a discontinuity in its
second derivative, i.e., in the specific heat. This phenomenon, first
hypothesized by Kauzmann, is called the ideal glass transition.  

\subsection{Zero temperature limit} The $q$-state anti-ferromagnetic
Potts model maps to a fundamental problem of graph theory: if the
ground state energy is zero, the graph is colorable by $q$
colors so that neighboring spins have different colors, otherwise the
graph is $q$-uncolorable. This has been studied in great
detail in
\cite{ColoringSaad,Coloring,ColoringFlo,ColoringMarc,MM05,US1,US2}.
An important relation is that if $m^*(T=0)>0$, then the ground state energy is zero and the graph is $q$-colorable. If
$m^*(T=0)=0$, however, the ground state energy is positive and the
curve $m^*(T)$ approaches the origin with a tangent $y=m/T$.

The limit $T\to 0$ at fixed ratio $m/T$ corresponds to the original
zero-temperature solution of the coloring problem
\cite{Coloring,ColoringFlo} and more generally to the so-called
"energetic" cavity method introduced in \cite{MP03}.  In the $m-T$
diagrams, all curves $m(T)$ starting at the origin $(0,0)$ have
therefore their counterpart in $y={\rm d}m(T)/{\rm d}(1/T)|_{T=0}$ in
the zero temperature formalism.  It is actually a non-trivial check of
our numerical results that the different curves $m(T)$ we found match
perfectly the corresponding slopes $y$ computed in
\cite{Coloring,ColoringFlo}.  This shows however that the zero
temperature "energetic" approach~\cite{MP03,Science,Coloring} does not
describe correctly the thermodynamical states in the colorable phase
(where $m^*>0$).  The results of this approach should thus be used
with caution, and have a clear meaning only in the $m-T$ phase
diagram.
This warning applies in particular to the critical values for the
dynamical and the 1RSB stability transitions in
\cite{BiroliMezard,Science,Coloring,ColoringFlo,MPR03}. A
complementary "entropic" zero temperature approach was however
introduced in \cite{ColoringMarc} and allows to obtain more
informations about the zero energy states \cite{US1,US2}.

\subsection{Gardner instability}
In the mean field spin glass theory, the 1RSB description may be wrong at low
temperatures, in which case more steps of RSB are necessary: this phenomena is
called the Gardner transition~\cite{Gardner,Sompolinsky}. To test the validity
of the 1RSB, it is widely considered as sufficient to perform a local
stability analysis of the 1RSB approach~\cite{MR03,MPR03,ColoringFlo,BiroliMezard}.
Two types of perturbation are to be considered. In the first type, one checks
if the 1RSB iterations (\ref{1RSB}) are stable towards small changes in the
site dependent distributions $P(\vec \psi)$.  In the second type, one checks
if the 1RSB iterations (\ref{1RSB}) are stable against small changes in the
probabilities $\vec \psi$ themselves.

The instability of the first type at positive temperature is quite problematic
to be checked. We used the analytical results of \cite{ColoringFlo} for the
$m,T \to 0$ limit with $y=m/T$ fixed (see the dotted lines $m_I$ in fig.~\ref{FIG-mt}). It is moreover a simple algebraic fact that the averages of the
1RSB distributions $P(\vec \psi)$ at $m=1$ follow the RS
equations~\cite{US1,US2}: for $m=1$ the 1RSB approach is thus type-I unstable
for $T<T_{\rm local}$. 
We conjecture that the contrary is also true, i.e. that non-convergence of the distribution $P(\vec \psi)$ at $m=1$ would manifest itself as a non-convergence of its mean. 

The instability of the second kind can also be studied analytically in the
$m,T \to 0$ limit~\cite{ColoringFlo}, and for generic $(m,T)$ numerically as
follows: one first find the fixed point of eq.~(\ref{1RSB}) using the
population dynamics method, and then compute how small differences in the
populations evolve in the iteration.  This can be done by creating a second
copy of the population representing the distribution $P(\vec \psi)$,
perturbing infinitesimally every of its elements and checking if the two
populations converged to the same one for long times, or equivalently by using
the more involved numerical method developed in \cite{MR03}, which we also
used to cross-check our results.

We observed that, at a given temperature, the 1RSB states
corresponding to $m>m_{I}$ are type I unstable while those
corresponding to $m<m_{II}$ are type II unstable. In order to check
the stability, one thus needs to compare $m_I(T)$ and $m_{II}(T)$ with
the thermodynamical $m^*(T)$. Generically, the instability of the
second type is the relevant one. Some examples are shown in
fig.~\ref{FIG-mt}: in (a) for $q=2$, $c=4$ the whole glassy phase is 1RSB
unstable.  In (b), $q=3$, $c=7$, the 1RSB solution is type II unstable
for $T<T_G$ and stable for $T>T_G$. We cannot rule out a type I
instability, yet we believe this instability unlikely, given the slope
of $m_{I}(T)$ at the origin. We also implemented the 2RSB equations
and were not able to find any non-trivial solutions in the type II
stable region. For (c) $q=4$, $c=11$ and (d) $q=5$, $c=14$ we observed that
the whole low temperature phase is 1RSB stable.

\subsection{Overlap distributions}
The overlap probability distribution function is a useful order
parameter in spin glasses~\cite{BOOK} and allows to characterize the
different phases. Consider two configurations $\alpha$ and $\beta$
chosen uniformly at random at given temperature.  To define their
overlap we use the matrix $Q^{\mu \nu}= \sum_{i=1}^N
\delta(\sigma^\alpha_{i},\mu)\delta(\sigma^\beta_{i},\nu)/N$ and an
example of the definition of an overlap which is invariant under
permutation of colors in one of the configurations is $Q=
\sum_{\mu,\nu=1}^q Q^{\mu \nu}$.  In the liquid phase, the equilibrium
distribution of the overlap $P(Q)$ is just a delta function.  Below the dynamic
transition, it is the same, and becomes non-trivial only for
$T<T_K$. In this case, the $P(Q)$ is made of two peaks, 
their relative weights are a monotonous function of the parameter
$m^*$. For $m^* \to 1$ the weight of the second peak is near to zero,
for $m^*\to 0$ there is only one dominating state and the weight of
the second peak is near to one.  For $T<T_G$ the distribution $P(Q)$
is continuous as in the fully connected Ising spin glass model
\cite{BOOK}. At zero temperature the distribution $P(Q)$ is nontrivial
only if $0<m^*(T=0)<1$ (i.e. in the colorable phase.  \footnote{The
  triviality of $P(Q)$ at zero temperature was predicted in \cite{KM}
  but the argumentation in \cite{KM} is valid only in the uncolorable
  phase.})

\section{The phase diagram}
\begin{figure*}[!t]
  \begin{center}
    \includegraphics[width=0.49\textwidth]{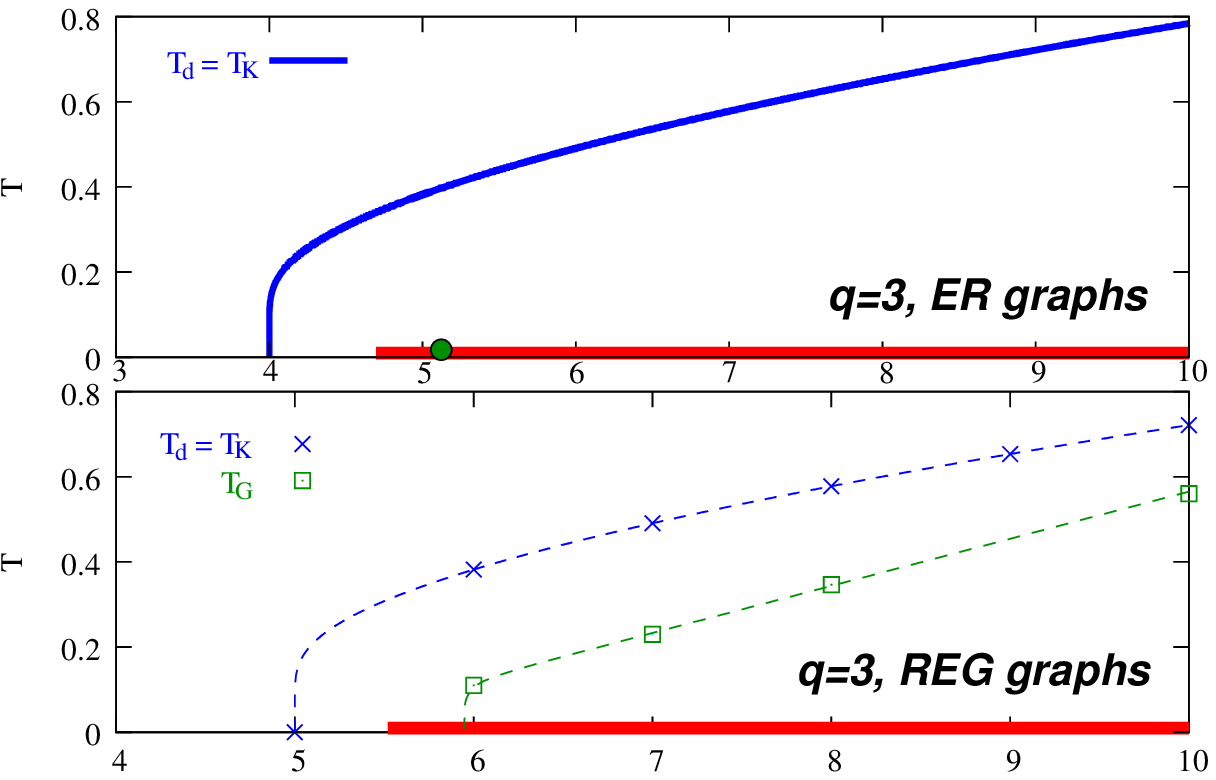}
    \includegraphics[width=0.49\textwidth]{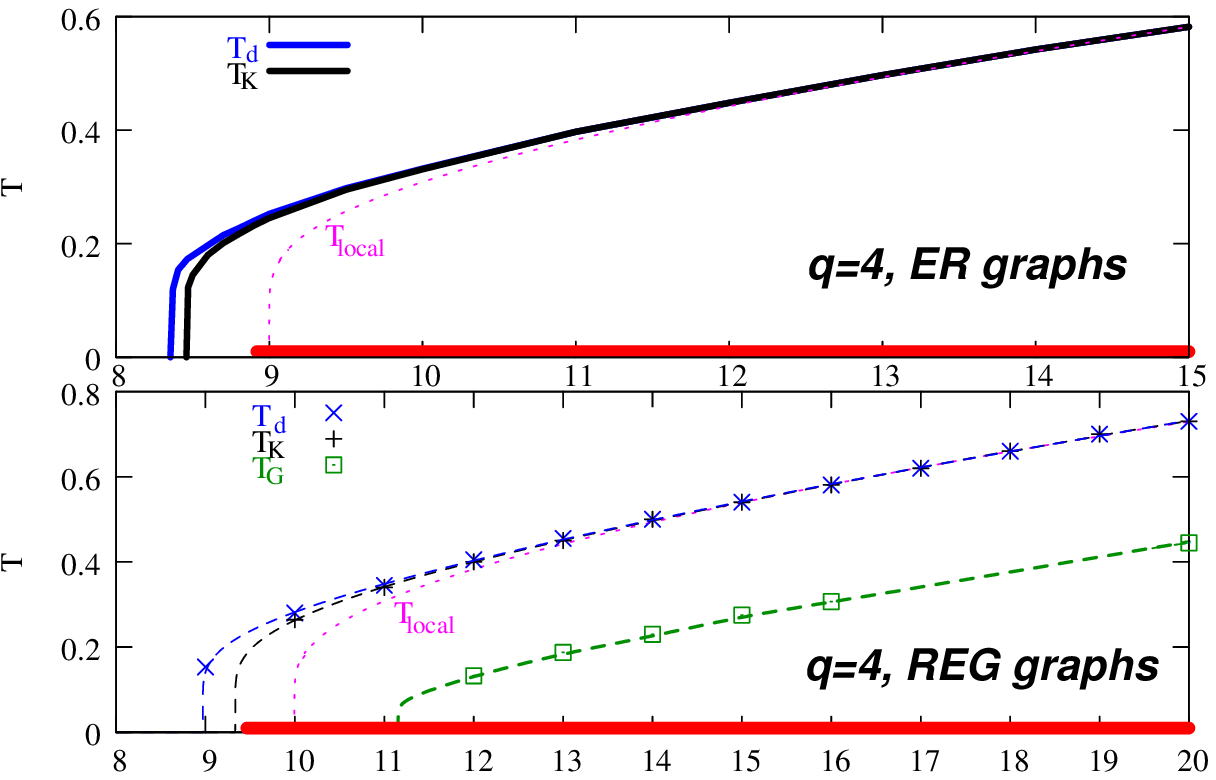}  
    \put(-370,-2){c}
    \put(-112,-2){c}
    \caption{Phase diagrams for the 3-state (left) and 4-state (right)
      anti-ferromagnetic Potts glass on Erd\H{o}s-R\'enyi graphs of
      average degree $c$ (top) and regular graphs of degree c
      (bottom). For $q=3$ the transition is continuous $T_d=T_K=T_{\rm
        local}$. For $q=4$, we find that $T_d > T_K > T_{\rm local}$,
      while for larger connectivities these three 
      temperatures become equal. The Gardner temperature $T_G$ for
      regular graphs is also shown (green online).  
      In the case of $q=3$ and ER graphs,
      $T_G>0$ only for connectivities larger than $c>c_G=5.08$ \cite{ColoringFlo}.
      The bold (red online) lines at zero temperature represent the uncolorable
      connectivities $c>c_s$.} 
\label{fig:Tc}
\end{center}
\end{figure*}
We now present our results and summarize our findings for the
temperature-connectivity phase diagrams.  We computed the dynamical $T_d$ and
Kauzmann $T_K$ temperatures both for regular and Erd\H{o}s-R\'enyi graphs via
the simplification of eq.~(\ref{1RSB}) at $m=1$ \cite{MM05}.  The Gardner
temperature $T_G$ is computed only in the simpler case of regular graphs.  

\subsection{Ising case: q=2} For the Ising spins, the spin glass
transition arises continuously at $T_{\rm local}=T_d=T_K$ given by eq.~(\ref{stab_KS}).  We found that the thermodynamic states given by
$m^*(T)$ are always 1RSB unstable as $m^*<m_{II}$ (see the example in
fig.~\ref{FIG-mt}(a)).  Ising spin glasses should thus be described
via the infinite replica symmetry breaking formalism as pointed out
already in \cite{MP99}.  We also notice that both the curves $m^*(T)$,
$m_{II}(T)$ both goes to zero when $T\to T^-_{\rm local}$, as in the
fully connected model.  In the zero temperature limit the random
graphs are $2$-uncolorable in the spin glass phase, i.e., when $T_{\rm
  local}>0$ (for $c>1$ on ER graphs and $c>2$ on regular graphs
\cite{ColoringSaad}).
\begin{table}[!t]
\begin{tabular}{|c||c|c|c|c|c||c|c|}\hline
  q & $ c $ & $T_{\rm d}$ & $T_{\rm K}$ & $T_{\rm local}$ & $T_{\rm G}$ & col \\
  \hline \hline
  2 & 3 & \multicolumn{4}{c||}{0.567} & no \\
  2 & 4 & \multicolumn{4}{c||}{0.759} & no \\
  \hline \hline
  3 & 5 & \multicolumn{3}{c|}{0} & xxx & yes \\  
  3 & 6 & \multicolumn{3}{c|}{0.381} & 0.11(1) & no \\  
  3 & 7 & \multicolumn{3}{c|}{0.490} & 0.23(1) & no \\  
  3 & 8 & \multicolumn{3}{c|}{0.577} & 0.34(2) & no \\  
  3 & 10 & \multicolumn{3}{c|}{0.721} & 0.56(3) & no \\  
  \hline 
  4 & 9 & 0.153(5)  & xxx & xxx & xxx & yes \\  
  4 & 10 & 0.280(5) & 0.264(5) & 0 & xxx & no \\
  4 & 11 & 0.345(5) & 0.340(5) & 0.308 & xxx & no \\
  4 & 12 & 0.405(5) & 0.400(5) & 0.382 & 0.135(5)& no \\
  4 & 13 & 0.455(5) & 0.450(5) & 0.441 & 0.185(5)& no \\
  4 & 14 & 0.500(5) & 0.500(5) & 0.492 & 0.230(5)& no \\
  4 & 15 & 0.540(5) & 0.540(5) & 0.539 & 0.275(5)& no \\
  4 & 16 & 0.581(5) & 0.581(5) & 0.581 & 0.305(5)& no \\
  4 & 20 & 0.730(5) & 0.730(5) & 0.730  & 0.445(5)& no \\
  \hline
  5 & 14 & 0.214(5) & 0.168(5) & xxx & xxx & yes \\  
  5 & 15 & 0.276(5) & 0.255(5) & xxx & xxx & no \\
  5 & 16 & 0.322(5) & 0.306(5) & xxx & xxx & no \\
  5 & 17 & 0.360(5) & 0.348(5) & 0 & xxx & no \\
  5 & 18 & 0.396(5) & 0.386(5) & 0.268 & 0.09(2) & no \\
  5 & 19 & 0.428(5) & 0.420(5) & 0.325 & 0.13(1) & no \\
  5 & 20 & 0.460(5) & 0.452(5) &  0.369 & 0.16(1) & no \\
  \hline 
\end{tabular}
\caption{Critical temperatures $T_d,T_K,T_{\rm local}$ and $T_G$  for the regular graphs ensemble and the $q$-colorability of the graphs. The error bars come from the numerical precision in evaluation the solution of (\ref{1RSB}) by the population dynamics method. }
\label{table}
\end{table}

\subsection{The case q=3} The phase diagrams are shown in
fig.~\ref{fig:Tc} (left). We again observe, for both ER and regular
graph, a continuous transition at $T_{\rm local}=T_d=T_K$ given by
eq.~(\ref{stab_KS}). However, the situation is different from the
Ising case. Consider for instance the case $c=7,q=3$ in
fig.~\ref{FIG-mt}(b): first, $m^*(T_{\rm local})=1$ and $m_{II}(T_{\rm
  local})=1$, and moreover the thermodynamical states are type II
stable for $T>T_G$.  Even if we have not ruled out completely the
instability of type I, this makes plausible the presence of a
continuous transition towards a 1RSB stable phase, just like in the
fully connected $3$-state Potts model of \cite{Sompolinsky}.

We found that the Gardner temperature is positive only in the uncolorable
phase. See table \ref{table} for critical values in regular graphs. 
For ER graphs it was already shown that $T_G>0$ only for $c>5.08$~\cite{ColoringFlo}. 
The zero temperature study~\cite{ColoringFlo} also indicates that both instabilities are
irrelevant close to the colorable threshold. All this suggests that, for colorable connectivities, the whole low temperature phase of the $3$-state Potts model is 1RSB stable.

\subsection{The case $q\ge 4$} In this generic case we observe the
same set of transitions as in the mean field theory for the ideal
glass transition in structural glasses.  We plot the $q=4$ phase
diagram in the right panel of fig.~\ref{fig:Tc}. The discontinuous
dynamical transition $T_d$ first arises followed by the Kauzmann
transition at $T_K<T_d$. The Gardner transition $T_G<T_K$
arises again for connectivities larger than the colorable
thresholds. It is also interesting to notice that as the connectivity $c$
grows, one actually observes that $T_d,T_K \to T_{\rm local}$ (see
table \ref{table}) so that a situation similar to $q=3$ is
recovered. This happens however for connectivities well beyond the coloring
transition $c_s$ (since $c_{s}\propto 2q \log{q}$ while $c_{\rm local}
\propto q^2$.).

Our results on the 1RSB type II stability, and the conjecture
that the system is type I stable for $T>T_{\rm local}$ at all
$m\le 1$, have important implications in the $q$-coloring problem: for
$q\ge 4$ the colorable phase is always 1RSB stable. The expert
reader will notice that the conclusions in \cite{ColoringFlo} were
different and that a unstable region was found,  \cite{ColoringFlo} however
describes the region $m,T\to 0$, which is not relevant for the {\it
  thermodynamical states} in the colorable phase. While {\it most}
clusters might be instable, the {\it relevant ones} are always
stable in the colorable phase.

\section{Discussion}
We studied in this paper a mean field Potts glass with
anti-ferromagnetic (or disordered) interactions. We observed a
continuous glass transition for $q=2,3$ and a discontinuous one for
$q\ge 4$ (at least for low enough connectivities).  We also considered
the stability of the 1RSB solution and concluded, in particular, that
the colorable phase at $T=0$ is stable for $q\ge4$ and probably for
$q=3$ as well, thus correcting previous claims\cite{ColoringFlo}.  We
expect this conclusion to be also valid for the satisfiability problem
\cite{MPR03}.
It is, however, important to develop a way to confirm properly the
type I stability for $q=3$.

The phase diagram for $q \geq 4$ (right panel of fig.~\ref{fig:Tc})
has many common characteristics with the one of the finite
connectivity p-spin model~\cite{ferro,MR03}, whose zero temperature
limit maps to the XOR-SAT problem~\cite{MRZ02}.  Let us, however,
state several differences: The satisfiability threshold $c_{s}$
corresponds to the Kauzmann one $c_K$ in \cite{ferro,MR03}, while they
are different in our model. The temperatures $T_K$ and $T_d$ are
different in the large connectivity limit in \cite{ferro,MR03} where
no local instability exists \footnote{Note that $T_{\rm local}=0$ in
  the fully connected p-spin, the RS transition being first order, and
  consistently no type I instability is found.}, while $T_K$ and $T_d$
converge to the same value $T_{\rm local}>0$ for large connectivities
in our model.  This illustrates the differences (and the richness) in
the phenomenology of the coloring and satisfiability problems when
compared to XOR-SAT.

Our model is also very similar to the lattice glass
model\cite{BiroliMezard} where the inverse temperature is replaced by
the chemical potential $\mu$ and the energy by the density of
particles $\rho$.  The close packing limit studied in
\cite{BiroliMezard} however concerns only the $m=0,\mu=\infty$ limit.
A quantitative study of the phase diagram with the correct value for
the dynamical, Kauzmann and Gardner transition in this model is thus
still missing and we are currently working in this direction.

We see many directions in which this work can be continued. First, it
would be interesting to consider the large connectivity limit for each
values of $q$ using the replica method, as was
done for 3-satisfiability in \cite{ParisiLeuzzi}. It would
also be of interest to study the dynamics along the line of
\cite{MR03}. Monte-Carlo simulations of these models (or exact ground
state enumerations) would also be valuable to cross-check the cavity results
and to study
finite-size effects. Finally, the disordered version of our model is a
good candidate for a glass former in finite dimension.

\begin{acknowledgments}
  This work has been partially supported by EVERGROW (EU consortium
  FP6 IST). We thank T. J\"org for a critical reading of the manuscript.
\end{acknowledgments}


\begin{thebibliography}{99}
\vspace{-0.4cm}  
\bibitem{SK} Sherrington D., Kirkpatrick S., {\it Phys. Rev. Lett.}
  {\bf 35}, 1792 (1975).

\bibitem{BOOK} Parisi G., {\it J. Phys.} {\bf A13} L115-L121, (1980).
  M\'ezard M., Parisi G., Virasoro M.A., {\it Spin Glass Theory and
    Beyond} (World Scientific, Singapore, 1987).

\bibitem{Talagrand} Talagrand M., {\it Annals of Math.} {\bf 163}, no
  1, 221-263 (2006).

\bibitem{Pspin} Gross D.J., M\'ezard M., {\it Nuclear Physics B},
  {\bf 240}, 4, 431-452 (1984). Derrida B., {\it Phys. Rev. Lett.}
  {\bf 35}, 1792 (1975).

\bibitem{Sompolinsky} Gross D.J., Kanter I., Sompolinsky H., {\it
    Phys. Rev. Lett.} {\bf 55}, 304 (1985). Kanter I., Sompolinsky H. 
  1987 {\it J. Phys. A} {\bf 20} L673. 

\bibitem{KirkpatrickThirumalai} Kirkpatrick T., Thirumalai D., {\it
    Phys. Rev.  Lett.} {\bf 58}, 2091 (1987). Kirkpatrick T.,
  Thirumalai D., Wolynes P., {\it Phys.  Rev. A} {\bf 40}, 1045
  (1989). Kirkpatrick T., Wolynes P., {\it Phys. Rev. A} {\bf 35},
  3072 (1987). 

\bibitem{BiroliMezard} Biroli G., M\'ezard M., {\it Phys. Rev. Lett.} {\bf 88}, 025501 (2001). 
 Rivoire O., Biroli G., Martin O.C., M\'ezard M., {\it    Eur. Phys. J. B} {\bf 37}, 55-78 (2004).

\bibitem{VianaBray} Viana L., Bray A.J., {\it J.~Phys.~C}
  \textbf{18}, 3037 (1985). Thouless D.J., {\it Phys. Rev. Lett.} {\bf
    56}, 1082 - 1085 (1986).

\bibitem{MP99} M\'ezard M., Parisi G., {\it Eur. Phys. J. B}
  {\bf 20}, 217 (2001).

\bibitem{MP03} M\'ezard M., Parisi G., {\it J. Stat. Phys} {\bf
    111} 1 (2003).

\bibitem{Science} M\'ezard M., Parisi G., Zecchina R., {\it
    Science} {\bf 297}, 812 (2002). M{\'e}zard M., Zecchina R.,
  {\it Phys. Rev. E} {\bf 66}, 056126 (2002).

\bibitem{Coloring} Mulet R., Pagnani A., Weigt M., Zecchina R.,
  {\it Phys. Rev. Lett.} {\bf 89}, 268701 (2002).  Braunstein A.,
  Mulet R., Pagnani A., Weigt M., Zecchina R., {\it Phys. Rev. E}
  {\bf 68}, 036702 (2003).

\bibitem{ColoringSaad} van Mourik J., Saad D., {\it Phys. Rev. E}
  {\bf 66}, 056120 (2002).

\bibitem{ColoringFlo} Krz\k{a}ka{\l}a F., Pagnani A., Weigt M.,
  {\it Phys. Rev. E} {\bf 70}, 046705 (2004).

\bibitem{ColoringMarc} M\'ezard M., Palassini M., Rivoire O.,
  {\it Phys. Rev.  Lett.} {\bf 95}, 200202 (2005).

\bibitem{MM05} M\'ezard M., Montanari A., {\it J. Stat. Phys.} {\bf
    124}, 1317 (2006).

\bibitem{US1} Krz\k{a}ka{\l}a F., Montanari A., Ricci-Tersenghi F., Semerjian
  G., Zdeborov\'a L., {\it Proc. Natl. Acad. Sci.} {\bf 104}, 10318
  (2007).

\bibitem{US2} Zdeborov\'a L., Krz\k{a}ka{\l}a F., {\it Phys. Rev. E} {\bf
    76}, 031131 (2007).

\bibitem{ElderfieldSherrington} Elderfield D., Sherington D.,
  {\it J. Phys. C} {\bf 16}, L497 (1983).

\bibitem{Enzo} Nishimori H., Stephen M.J., {\it Phys. Rev. B} {\bf
    27}, 5644 - 5652 (1983); Marinari E., Mossa S. and Parisi G., {\it
    Phys. Rev. B} {\bf 59}, 8401-8404 (1999).

\bibitem{Glass} Montanari A., Semerjian G., {\it J. Stat. Phys.} {\bf
    124}, 103 (2006).
 
\bibitem{Remi} Monasson R., {\it Phys. Rev. Lett.} {\bf 75}, 2847 (1995).

\bibitem{MPR03} Montanari A., Parisi G., Ricci-Tersenghi F., {\it
    J.  Phys. A} {\bf 37}, 2073 (2004); Mertens S., M\'ezard M., Zecchina
  R., {\it Rand. Struct. and Algo.} {\bf 28}, 340 (2006).

\bibitem{Gardner} Gardner E., {\it Nucl. Phys. B} {\bf 297}, [FS14] 74
  (1985).

\bibitem{MR03} Montanari A., Ricci-Tersenghi F., {\it Eur. Phys. J. B}
  {\bf 33}, 339 (2003).

\bibitem{KM} Krz\k{a}ka{\l}a F., Martin O.C., {\it Europhys. Lett.}
  {\bf 6}, 749 (2001).

\bibitem{ferro} Franz S., M\'ezard M., Ricci-Tersenghi F., Weigt M.,
  Zecchina R., {\it Europhys. Lett.} {\bf 55}, 465 (2001).

\bibitem{MRZ02} M\'ezard M., Ricci-Tersenghi F., Zecchina R., {\it
    J. Stat. Phys.} {\bf 111}, 505 (2002).

\bibitem{ParisiLeuzzi} Leuzzi L., Parisi G., {\it J Stat Phys} {\bf
    103} 679 (2001); Crisanti A., Leuzzi L., Parisi G., {\it
    J. Phys. A} {\bf 35}, 481 (2002).

\end{thebibliography}
\end{document}